# Unveiling tortured phrases in Humanities and Social Sciences

Alexandre Clausse[1], Fidan Badalova[2], Guillaume Cabanac[1,3], and Philipp Mayr[2]

[1] *{first.last}@univ-tlse3.fr*
Université de Toulouse, IRIT UMR 5505 CNRS, 118 route de Narbonne, 31400 Toulouse (France)

[2] *{first.last}@gesis.org*
GESIS - Leibniz Institute for the Social Sciences, Unter Sachsenhausen 6-8, 50667 Köln (Germany)

[3] Institut Universitaire de France (IUF), Paris (France)

## Abstract

A small amount of unscrupulous people, concerned by their career prospects, resort to paper mill services to publish articles in renowned journals and conference proceedings. These include patchworks of synonymized contents using paraphrasing tools, featuring tortured phrases, increasingly polluting the scientific literature. The Problematic Paper Screener (PPS) has been developed to allow articles (re)assessment on PubPeer. Since most of the known tortured phrases are found in publications in science, technology, engineering, and mathematics (STEM), we extend this work by exploring their presence in the humanities and social sciences (HSS). To do so, we used the PPS to look for tortured abbreviations, generated from the two social science thesauri ELSST and THESOZ. We also used two case studies to find new tortured abbreviations, by screening the Hindawi EDRI journal and the GESIS SSOAR repository. We found a total of 32 multidisciplinary problematic documents, related to Education, Psychology, and Economics. We also generated 121 new fingerprints to be added to the PPS. These articles and future screening have to be investigated by social scientists, as most of it is currently done by STEM domain experts.

## Introduction

Scientific research is a cumulative process involving rigorous reporting of the experiments carried out and the observed results, in a textual or visual form, typically presented as scientific articles. These findings are then submitted to an editorial board or group of peers in order to be published, after a peer review process. The so-called 'publish or perish' paradigm implies to publish as many articles as possible in reputable journals to have the most impactful research in a given scientific community (Biagioli & Lippman, 2020). The publication pressure on individual researchers leads a small number of unscrupulous people, concerned about their career prospects, to resort to falsification, fabrication, and plagiarism.

Plagiarism can be disguised using paraphrasing tools such as spinners (e.g., SpinBot) to rephrase textual contents, including established scientific concepts. When a scientific concept is paraphrased, at least one of its terms gets replaced by a synonym, making it nonsensical regarding the associated discipline. For example, a 'convolutional brain organization' is a spun version of the 'convolutional neural network', which is known as a tortured phrase (Cabanac, Labbé & Magazinov, 2021). They are assumed to be evidence of paper mill products, a company that sells fake scientific articles, ensuring that they will be published in known journals, by manipulating editorial and publishing processes (Abalkina *et al.*, 2025; Nazarovets, 2024). The consequence of such behavior is twofold: (1) some (sensitive) research relies on these articles, making them unreliable, and (2) this leads to a major trust issue in science.

To address this, the Problematic Paper Screener (PPS) was launched in 2021; it builds upon the Dimensions bibliometric database full-text search, and allows (re)assessing questionable articles on the PubPeer platform (Cabanac, Labbé & Magazinov, 2021; Barbour & Stell, 2020). As of today, the 'tortured' detector flagged more than 18k scientific articles containing at least 5 different tortured phrases, only 2.9k of which have been retracted. However, this is tedious work, requiring several domain experts to read each article to update the PPS fingerprints list (i.e., known tortured phrases). Moreover, not all the disciplines have yet been considered, as they are mostly related to science, technology, engineering, and mathematics (STEM) studies.

We propose to extend this work and search for the presence of tortured phrases in humanities and social sciences (HSS) articles and developing guidelines to raise the awareness of the scientific community regarding such fraudulent content.

**Motivation**

Tortured phrases are widely polluting the scientific literature (Van Noorden, 2023). Some of them are claiming false information (Texeira da Silva, 2021), even in sensitive research related to COVID-19 (Texeira da Silva, 2023). These publications are unreliable and causing major trust issues in science. Moreover, some of these articles are used as the foundations for other studies, and false information and errors spreads. These articles featuring unreliable references are known as 'feet of clay' publications (Cabanac, 2024).

In 2022, a post publication peer-review (PPPR) approach using the Problematic Paper Screener (PPS) and PubPeer has been proposed as part of an initiative to decontaminate the scientific literature (Cabanac, 2022). It focuses on two main tasks: (1) investigate the suspect paper and (A, B) extract all the problematic content to check if (C) it has been commented on PubPeer, and (2) (re)assess it using (A) the PPS and (B) PubPeer in order to discuss its content, as depicted in Figure 1.

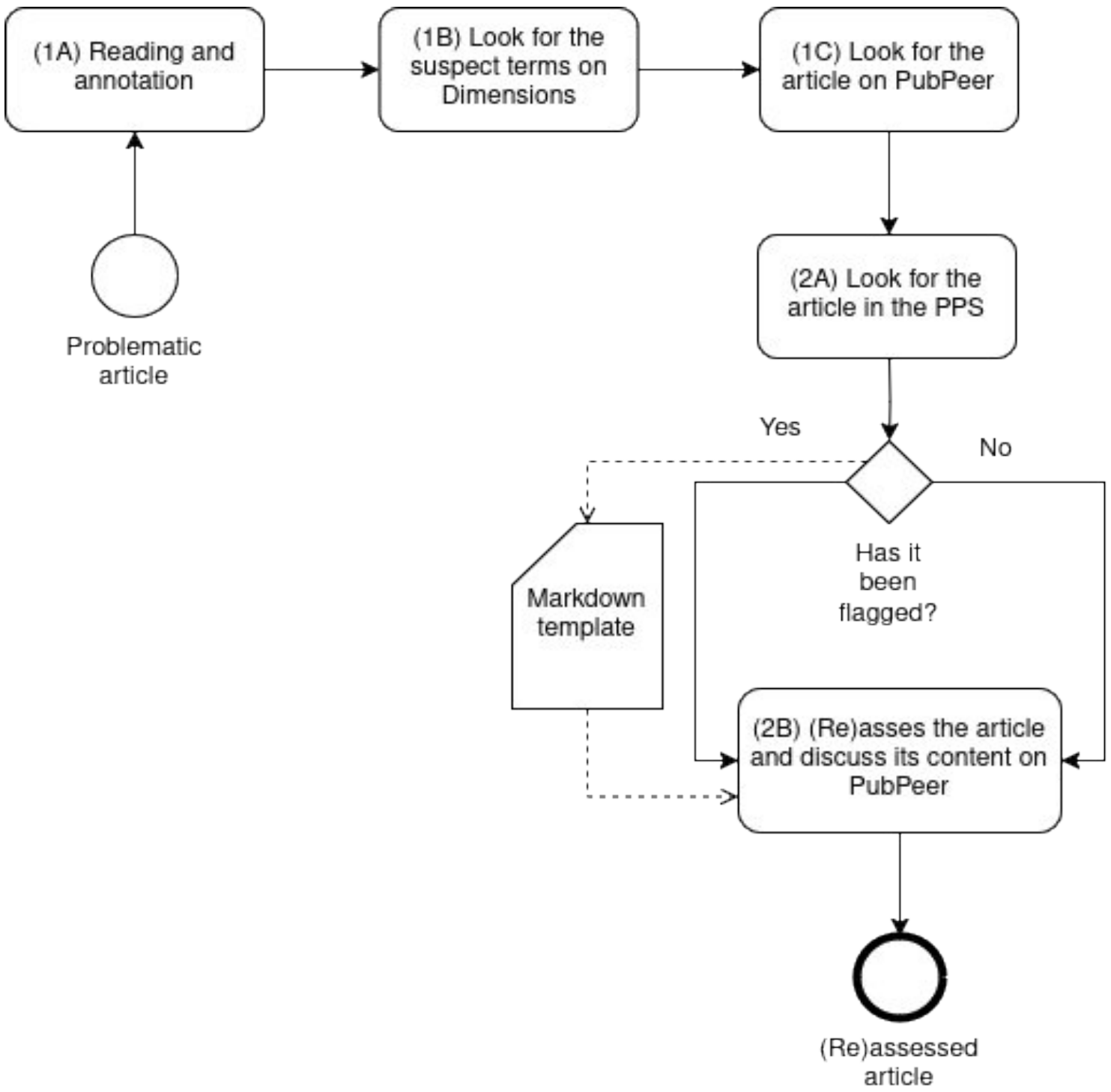

**Figure 1. The research quality insurance workflow, describing how an analyst looks for tortured phrases in an article flagged by the PPS and (re)assesses it on PubPeer.**

As of today, the PPS features 6,545 fingerprints (i.e., known tortured phrases) which are used to screen the scientific literature by querying the Dimensions database, as their documents textual contents are indexed in. Since the 'tortured' detector already found tortured phrases in 18,582 document, most of them are related to STEM studies, as depicted in Figure 2.

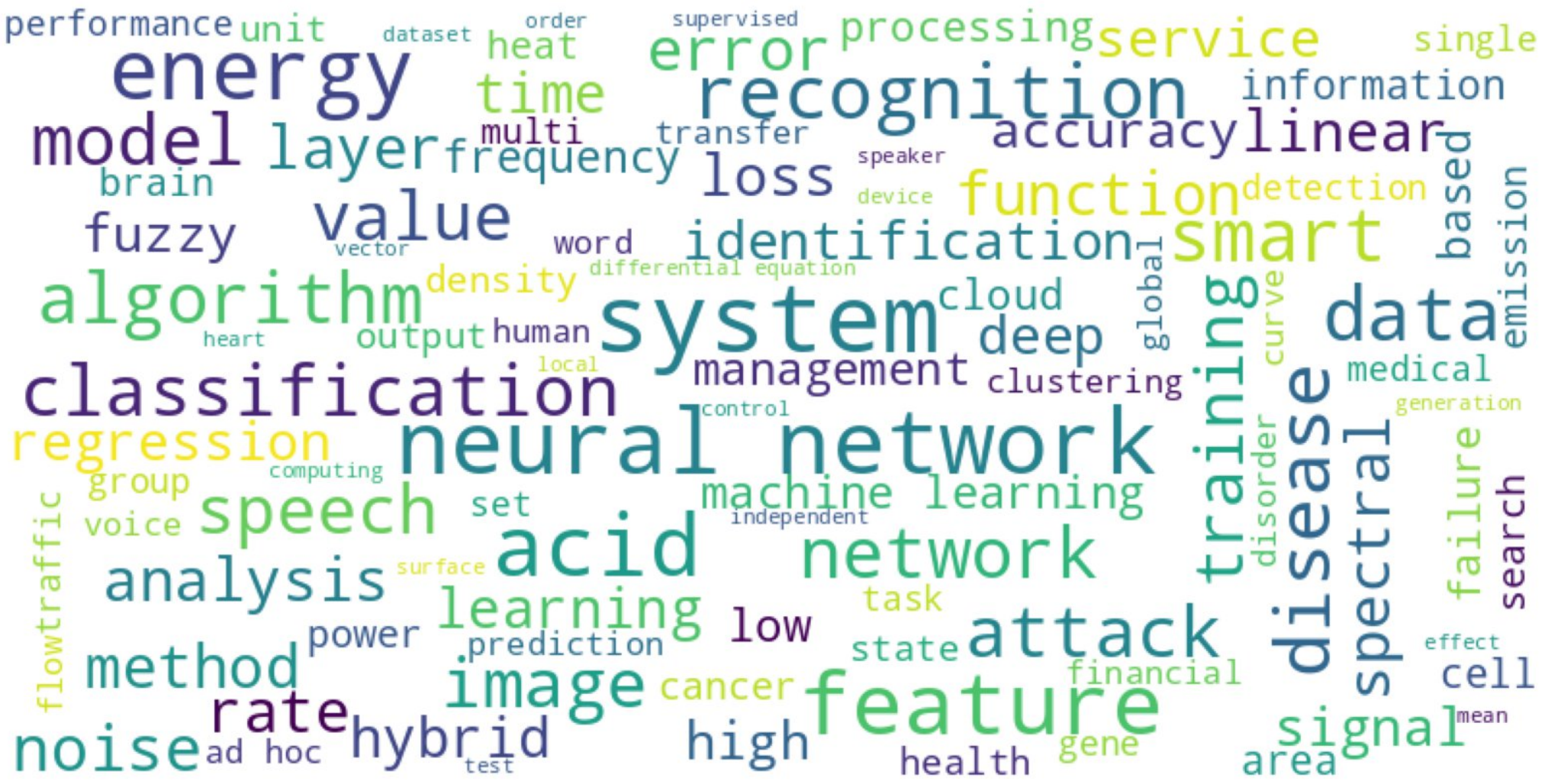

**Figure 2. Word cloud of terms and phrases from the fingerprints list.**

However, exploring further this list of fingerprints, it appears that it also contains 42 tortured phrases related to HSS (e.g., 'electronic democratic framework', the tortured phrase for 'electronic voting machine'). Moreover, we found irregularities while looking at the term 'computer-assisted telephone interviewing (CATI)', where the word 'telephone' has been replaced by 'phone' in several scientific articles, making the abbreviations mismatching its developed form.

We studied the European Language Social Science Thesaurus (ELSST) and the GESIS Thesaurus for the Social Sciences (THESOZ) thesauri that showed subtleties to be taken into account. As an example, 'civil war' and 'internal war' are genuine concepts, although we first thought one was the tortured version of the other. This suggests that the HSS vocabulary is less standardized as the STEM one.

## Materials and method

We adopted a systematic approach to identify bogus text by focusing on tortured abbreviations (Clausse, 2023; O'Grady, 2024): tortured phrases mismatching their abbreviations (e.g., 'World Exchange Association (WTO)', the tortured version of 'World Trade Organization'). This approach is part of the Tortured Phrases ToolKit (TPTK) initiative, following the above presented research quality insurance workflow. To do so, we extracted all the abbreviations contained in the ELSST ($n = 60$) and THESOZ ($n = 75$) thesauri, we spun them using SpinBot, then we filtered out the unaltered abbreviations, ending with a total of 121 tortured abbreviations (Clausse *et al.*, 2025).

We included these generated tortured abbreviations in the PPS fingerprints list, then screened the scientific literature. We cross-validated these outputs using Dimensions, to filter out STEM-only documents according to the Australian and New Zealand Standard Research Classification (ANZRC) 2020 standards. We excluded paywalled articles, erroneous landing pages, defunct DOIs, and contents not featuring any tortured abbreviation.

We tested our approach on two case studies to screen articles included in the Hindawi Education Research International (EDRI) journal, and documents indexed by the GESIS Social Science Open Access Repository (SSOAR). We chose to do so as the open access publisher Hindawi, which is now owned by Wiley, has been the victim of a large-scale manipulation, leading to the publication of many paper-mill articles featuring tortured phrases. They released a full XML dump of their publications, which is unfortunately unavailable since June 2024. The EDRI journal contains a total of 760 articles. We also explored the GESIS SSOAR repository to ensure that no fraudulent content have been yet indexed. As of today, it contains 87,233 documents.

Thus, we used both the 42 tortured phrases and 121 tortured abbreviations (Clausse *et al.*, 2025) to explore the documents indexed by Dimensions, contained in the Hindawi EDRI journal, and indexed by the GESIS SSOAR (see Table 1). We lately extended the last exploration by mining all the abbreviations contained in the documents indexed in the latter repository, to find potentially new tortured abbreviations, and evaluate the generalization of the TPTK software.

**Table 1. Examples of tortured phrases and tortured abbreviations related to HSS, used as fingerprints to flag problematic articles.**

| *Tortured phrase* | *Expected term* |
|---|---|
| Academic substantive information (PCK) | Pedagogical content knowledge (PCK) |
| Non-administrative associations (NGOs) | Non-governmental organizations (NGOs) |
| Communities for infectious prevention and anticipation (CDC) | Centers for disease control and prevention (CDC) |
| Uprightness of the votes | Electoral integrity |
| Trickery in conduct | Fraud |
| Geological locale | Geographical locations |

## Results and discussion

Exploring the PPS, we matched 543 documents featuring at least one of the 121 generated tortured abbreviations, and 107 additional documents matching at least one of the already referenced 42 HSS fingerprints. After filtering out the irrelevant articles and assessing the remaining ones, we found a total of **26 problematic documents**. We found between 1 and 6 distinct fingerprints in each document, these have been published between 2017 and 2024. They are either preprints from the Social Science Research Network (SSRN), then articles and proceedings from both local institutions and the 'haute couture' of scientific literature (such as Elsevier, IEEE, and Wiley).

Exploring the Hindawi EDRI journal, we found one article published in 2021, featuring 5 tortured abbreviations. Some of them were not yet part of the PPS fingerprint list. Following the same approach, we did not find any problematic document in the GESIS SSOAR repository. Finally, finding all the possible abbreviations in the GESIS SSOAR documents yielded a total of 23,477 abbreviations including 9,322 labelled as 'tortured', as depicted in Table 2.

**Table 2. Processed documents from the GESIS SSOAR repository.**

| *Type* | *Count* |
|---|---|
| Total documents | 87,233 |
| English documents | 33,748 |
| Documents featuring abbreviations | 23,477 |
| Documents featuring tortured abbreviations | 9,322 |
| Validated false positives | 5,048 |

We manually validated these results (as of today, we checked 5,048 of them), and the majority of them were false positives, such as foreign institutions (e.g., ‘National Centre for Scientific Research (CNRS)’ for the ‘Centre National de la Recherche Scientifique’) and reversed words (e.g., ‘Hypothesis of Rational Expectations (REH)’ instead of ‘Rational Expectation Hypothesis’), making them genuine. However, we found 5 more problematic documents to be (re)assessed. These are preprints, articles, and monographs published between 2023 and 2024.

These 32 flagged documents are multidisciplinary and related to Education, Psychology, and Economics. Their screening highlighted new filtering rules to be implemented through the TPTK tortured abbreviations detector. As a contribution, we made new comments on the PubPeer platform to (re)assess 4 documents[1], which contain at least 4 distinct tortured abbreviations. We are aware that some of the matched abbreviations may still be false positives as they may have different meanings given the HSS field of research, and since they are less normalized as for STEM studies.

## Conclusion

In this study, we explored the presence of tortured phrases in HSS articles. Using SpinBot, we generated 121 tortured abbreviations to be included in the PPS fingerprints list, in addition to the 42 HSS tortured phrases already referenced. We flagged a total of 32 multidisciplinary documents featuring tortured abbreviations related to Education, Psychology, and Economics, however we could not process the closed access ones since they are behind paywalls. We also found new filtering rules to be implemented through TPTK, to improve the precision of this software.

So far, we made 4 new comments on the PubPeer platform to alert readers that tortured phrases are also features in HSS articles. These flagged publications should be investigated by social scientists, as the domain experts working on the PPS are mostly related to STEM, and the HSS vocabulary is less normalized. However, more than just being aware of inconsistencies, actions should be taken against the articles assessed as fraudulent, by retracting them.

Finally, we proposed guidelines to encourage the scientific community to be aware of such fraudulent content, as a research quality insurance. We invite anyone interested in reassessing fraudulent articles to take part of the decontamination of the scientific literature, as an opportunity to embrace an established method.

## Acknowledgments

We thank the political scientist Guillaume Levrier, who has been flagging several tortured phrases[2] in social sciences articles we could use in our study. We also would like to thank all the people who devote time and energy to decontaminate the scientific literature, mostly *pro bono*.
This article has benefited from an invited research stay at GESIS – Leibniz Institute for the Social Sciences, Köln, Germany via the Junior Research Call JRC-2024-02.
Alexandre Clausse and Guillaume Cabanac acknowledge the NanoBubbles project, that has received Synergy grant funding from the European Research Council (ERC), as part of the European Union's Horizon 2020 program, grant agreement number 951393. Fidan Badalova

1 https://pubpeer.com/search?q=%22several+tortured+abbreviations%22
2 https://pubpeer.com/search?q=%22Guillaume+Levrier%22

and Philipp Mayr received funding support from Deutsche Forschungsgemeinschaft (DFG), grant number MA 3964/15-3.